\documentclass[11pt]{article}
\usepackage{psfig,epsfig}

\newcommand{\refeq}[1]{(\ref{#1})}
\newcommand{\equ}[1]{(\protect\ref{#1})}

\newcommand{\al}{\alpha}
\newcommand{\la}{\lambda}
\newcommand{\ep}{\varepsilon}

\newcommand{\ptime}[1]{\frac{\partial #1}{\partial t}}
\newcommand{\ppar}{\nabla_\parallel}

\newcommand{\nuperp}{\nu_\perp}
\newcommand{\nupar}{\nu_\parallel}

\newcommand{\xpar}{x_\parallel}
\newcommand{\eh}{{\bf e}_h}
\newcommand{\eperp}{{\bf e}_\perp}
\newcommand{\epar}{{\bf e}_\parallel}
\newcommand{\Per}{\perp} \newcommand{\Par}{\parallel}
\newcommand{\reals}{{\rm I \mkern-2.5mu \nonscript\mkern-.5mu R}}
\renewcommand{\vec}[1]{{\bf #1}}
\newcommand{\FigPSPath}{.}

\begin{document}

\begin{titlepage}
\vspace*{1cm}
\begin{center}
  {\huge SCALING OF A SLOPE:\\\ \\
    THE EROSION OF TILTED LANDSCAPES\footnote{To appear in
    {\em Journal of Statistical  Physics}}}\\ 
  \vspace{2cm}
  {\Large Romualdo Pastor-Satorras and Daniel H. Rothman}\\
  \vspace{1cm} Department of Earth, Atmospheric, and Planetary
  Sciences\\
  Massachusetts Institute of Technology, Cambridge, Massachusetts
  02139 \vspace{2cm}

\abstract{ We formulate a stochastic equation to model the erosion of
  a surface with fixed inclination. Because the inclination imposes a
  preferred direction for material transport, the problem is
  intrinsically anisotropic.  At zeroth order, the anisotropy
  manifests itself in a linear equation that predicts that the
  prefactor of the surface height-height correlations depends on
  direction.  The first higher-order nonlinear contribution from the
  anisotropy is studied by applying the dynamic renormalization group.
  Assuming an inhomogeneous distribution of soil substrate that is
  modeled by a source of static noise, we estimate the scaling
  exponents at first order in $\ep$-expansion.  These exponents also
  depend on direction.  We compare these predictions with empirical
  measurements made from real landscapes and find good agreement.  We
  propose that our anisotropic theory applies principally to small
  scales and that a previously proposed isotropic theory applies
  principally to larger scales.  Lastly, by considering our model as a
  transport equation for a driven diffusive system, we construct
  scaling arguments for the size distribution of erosion ``events'' or
  ``avalanches.''  We derive a relationship between the exponents
  characterizing the surface anisotropy and the avalanche size
  distribution, and indicate how this result may be used to interpret
  previous findings of power-law size distributions in real submarine
  avalanches.}\\
\end{center}

\vspace*{\fill} {\bf Keywords:} Erosion, anisotropy, stochastic
equation, renormalization group

\end{titlepage}

\section{Introduction}

One of the great challenges faced by modern research in nonlinear
physics is the construction of predictive theories for systems in
which the underlying equations of motion are not known.  An example
that has recently created much interest is the flow of sand, or, more
generally, granular fluids \cite{jaeger96,kadanoff98}.  One expects
that the Navier-Stokes equations do not apply to granular flow because
of the dissipative nature of grain-grain collisions.  Although
physical arguments have been used to deduce equations of motion
applicable to certain situations (e.g., see Ref.~\cite{haff83}), a
comprehensive theory of granular flow remains elusive.

Still more complicated than sand is the 
flow of {\em wet} sand \cite{sand}.
Viewed naively, in such a situation we retain the complexity of the
Navier-Stokes equations and add to it the complicated frictional
stresses of a granular heap.  In this paper we shall concern ourselves
with wet sand in the form of a particular problem in geomorphology.
Specifically, we study the erosion of a landscape due, usually but not
exclusively, to the flow of water over it.  Despite the obvious
difficulties of this problem, our aim is to obtain some simple results
that offer fundamental insight into erosion.  Why we expect to do so
deserves some further comment.

There is now a wealth of empirical evidence that shows that real
landscapes exhibit some form of scale invariance \cite{iturbe97}.
These scaling laws come in many forms.  Perhaps best known are those
that describe the branching of river networks \cite{iturbe97}.  These
are not the scaling laws that interest us here, however.  Instead, we
wish to view the problem of topography more generally, by examining
statistical properties of surfaces $h({\bf x})$, where ${\bf x}$
denotes the horizontal position and $h$ is the topographic height or
elevation.

The statistics that primarily interest here are the height-height
correlation functions
\begin{equation}
  C ( {\bf r} )=\langle | h( {\bf x} + {\bf r}  ) - h( {\bf x} ) |^2
  \rangle_{\bf x}^{1/2}. 
\end{equation}
Scale invariance comes in the form of {\em self-affinity}
\cite{mandelbrot82}.  In other words, $C(r) \sim r^\alpha$, where
$\alpha$ is known as the {\em roughness exponent}.  Various empirical
measurements in different sorts of terrain show that this power-law
form may hold over an order of magnitude or more, with some
measurements indicating that $\alpha$ is small $(0.30 < \alpha <
0.55)$
[7--13]
while others show it to be large $(0.70 < \alpha < 0.85)$
[9--16].

Although there is not much agreement on the value of the scaling
exponent $\alpha$, the occurrence of scaling itself is fairly common.  
We are therefore led to consider theoretical models
whose solutions also exhibit scaling. 
These models are stochastic partial differential equations
that are Langevin equations for the evolution of a
surface \cite{halpin95}.  
One of our goals here is to derive such an equation that
predicts some aspects of the observed scaling.  Our hope is that our
predictions are general and independent of details such as material
properties, climate, etc.  Thus we hope that our model exhibits some
degree of {\em universality} \cite{kadanoff90}.

For our model to exhibit universality, we must identify a class of
topographic evolution problems for which we may make quantitative
predictions.  The class of problems we discuss here are problems in
which symmetry is broken by the existence of a preferred
direction---downhill---for the flux of eroded material.  Following
work we have already reported in a brief Letter \cite{pastor98}, we
derive an anisotropic noisy diffusion equation to describe erosion at
the small length scales where the preferred direction is fixed
throughout space.  The linear regime of this equation predicts that $C
( {\bf r} )$ is anisotropic at the level of its prefactors.  The
predicted anisotropy is testable, and empirical studies in progress
show that it works with unusual generality \cite{chan98}.  Under the
additional assumptions that the flux of eroded material increases with
increasing distance downslope and that the dominant effects of noise
are fixed in space, we find, using the dynamic renormalization group,
that not only is $C ( {\bf r} )$ anisotropic, but that it scales with
different exponents that correspond to the downhill direction and the
direction perpendicular to the downhill direction.  This result is
also testable, and we present one example, made from the topography of
the continental slope off the coast of Oregon, in good agreement with
our predictions.

An additional conclusion of our study concerns the wide range of
values of $\alpha$ that have been reported in the literature.  It has
been proposed previously \cite{sornette93} that observations of
$\alpha \simeq 0.4$ could be explained by the Kardar-Parisi-Zhang
(KPZ) equation \cite{kardar86}
\begin{equation}
  \label{kpz}
  \ptime{h} = \nu \nabla^2 h + \frac{\lambda}{2} | \nabla h |^2 + \eta .
\end{equation}
In its application to geomorphology, $\nu$ is a topographic
diffusivity coefficient, $\lambda$ is related to the velocity of
erosion in the direction normal to the surface, and $\eta$ is a source
of random noise that is uncorrelated in space and time.  We observe
here that if equation (\ref{kpz}) really does capture some aspects of
topographic evolution, then it applies only to those cases in which
$\alpha$ is found to be small.  It turns out that most observations of
small $\alpha$ are made at large length scales where no preferred
direction is easily identified
[7--13],
whereas observations of large $\alpha$ are usually associated with
small length scales
[9--16].
Because the average results predicted by our anisotropic theory are
consistent with these large-$\alpha$ observations, we can tentatively
identify two ``universality classes'' of topographic evolution.  In
the KPZ class, topography evolves isotropically (perhaps due to
internal tectonic stresses) at large length scales and yields small
roughness exponents, while in our anisotropic class, topography
evolves erosively, ``one slope at a time,'' at small length scales and
yields large roughness exponents.  We lend some support to this
conclusion by showing evidence of such a crossover in a single
topographic dataset.

The general framework of our theory also allows us to make some
contact with the larger field of ``self-organized criticality'' (SOC)
\cite{jensen98}.  Specifically, sloping submarine topography gives rise
to underwater avalanches.  These avalanches create flows, known as
turbidity currents \cite{simpson97}, that eventually come to rest as
sedimentary deposits called {\em turbidites} \cite{press82}.  A number
of recent studies have indicated that the size distribution of these
natural avalanches may follow the power-law scaling predicted by SOC
sandpile models \cite{hiscott92,rothman94,rothman95}.  Here we show
how our theory for topographic evolution may be linked to the SOC
theory for avalanche sizes.  Specifically, we derive a relation
between the anisotropic correlations of the slope and the size
distribution of the avalanches.

This long introduction will have succeeded if the reader is convinced
that the concepts of scaling and universality may have some
applicability to understanding some generic features of our natural
environment.  In this spirit, it is our pleasure to dedicate this
paper to Leo Kadanoff.

\section{Mathematical formulation}

We begin our discussion with a brief introduction to stochastic
equations for surface growth \cite{marsili96,barabasi95}.  We first
review some standard isotropic models, and briefly remark on their
applicability to geomorphology.  We then introduce our anisotropic
model.  Analysis of the model is deferred to the following section.

\subsection{Isotropic surface growth}

Our objective is to determine the evolution of the surface $h(\vec{x},
t)$, where, as we have already stated, $h$ is the height of the
surface at position $\vec{x}$ and time $t$.  We assume that $h$ is
single valued---that is, overhangs are not allowed.
The general form
of an equation for $h$ that we consider here is
\begin{equation}
  \label{hgen}
  \frac{\partial h(\vec{x}, t)}{\partial t} = {\cal F}[h(\vec{x}, t)] +
  \eta(\vec{x}, t).
\end{equation}
${\cal F}$ represents the flux of eroded material, and $\eta$ is a
source of random noise that allows us to include random fluctuations
in the growth process.  In the absence of specific information on
${\cal F}$, one generally seeks to first identify all applicable
physical symmetries and conservation laws.  This then allows the
construction of the simplest possible form of ${\cal F}$ compatible
with these constraints \cite{barabasi95}.

Of the vast number of equations such as \refeq{hgen} that have been
proposed in recent years \cite{marsili96,barabasi95}, here we restrict
the discussion to models that have been proposed for the study of
erosion at large length and long time scales.  These may be divided
roughly into two categories: those which conserve a material flux $\bf
J$ and those which do not.  In the conservative models, ${\cal F} = -
\nabla \cdot {\bf J}$, where ${\bf J}$ is the current of material.
The simplest of these has ${\bf J} = - \nabla h$ and no noise, leading
to the classical diffusion equation, which, in geomorphology, was
first popularized by Culling \cite{culling60}:
\begin{equation}
  \label{deq}
  \frac{\partial h}{\partial t} = \nabla^2 h.
\end{equation}
The diffusion equation alone, however, cannot explain the formation of
stable self-affine landscapes.  If we add uncorrelated noise, we
obtain the so-called Edwards-Wilkinson equation
\cite{whittle62,edwards82}.  With this equation we can obtain true
self-affine surfaces, but in the relevant number of dimensions (i.e.,
when the dimensionality of the position vector $\bf x$ is $d=2$), the
noisy version of equation \refeq{deq} predicts that correlations decay
logarithmically (i.e., $\al=0$) \cite{whittle62,edwards82}.  Thus
neither the deterministic nor stochastic form of equation \refeq{deq}
is compatible with the aforementioned observational evidence.

Partly as a remedy for this problem, non-conservative equations have
been proposed.  The most popular of these is the KPZ equation
\refeq{kpz}.  As Sornette and Zhang \cite{sornette93} have remarked,
the KPZ equation is attractive as a model of erosion because it is the
simplest surface growth equation that can generate a nontrivial
($\alpha \neq 0$) self-affine landscape.  Specifically, the KPZ
equation contains the necessary ingredients of nonlinearity and noise.
Roughly speaking, the nonlinearity results from a uniform rate of
erosion, at all locations $\bf x$, in the direction normal to the
surface, and the noise accounts for the irregularities of the process
in time and space.  The roughness exponent reported for KPZ in $d=2$
by numerical integration varies between $0.2$ and $0.4$
\cite{chakrabarti89,amar90,grossmann91}.  As discussed above, these
values are in reasonable agreement with observations at large
length scales, where one finds that $0.30 < \alpha < 0.55$
[7--13].
Thus the KPZ scenario of non-conservative isotropic growth normal to
the surface may indeed apply to some aspects of large-scale landscape
evolution.

\subsection{An anisotropic model}

\begin{figure}[t]
  \centerline{\epsfig{file=\FigPSPath/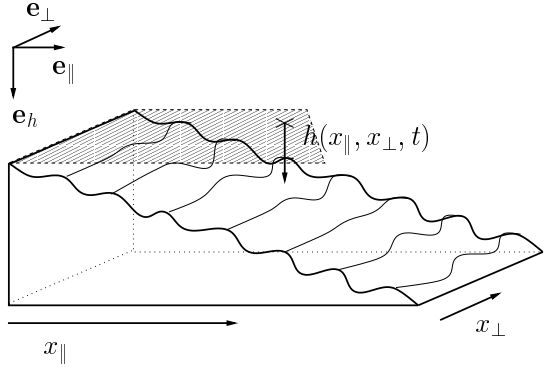, width=8truecm}}
  \caption{Schematic configuration of an anisotropic landscape in
    $d=2$.}
  \label{hillslope}
\end{figure}

As discussed in the introduction, the predictions of the KPZ equation
do not agree with many measurements made on landscapes at small length
scales
[9--16].
Thus some other physical mechanisms must be dominant in this range.
Here we propose that evolution at small length scales is strongly
influenced by the breaking of symmetry induced by the presence of a
slope of fixed inclination.

Figure~\ref{hillslope} depicts the framework for our theory.  The
vector $\eh$ is the growth direction in our parametrization.  Note
that $\eh$ is measured from the {\em top} of the slope downwards.  The
action of gravity selects a preferred direction given by the vector
$\epar$, which is essentially defined by the direction ``downhill,''
and which coincides with the average direction of the flow of
material.  The symbol $\eperp$ stands in general for the subspace of
directions perpendicular to $\epar$. For a real landscape, this
subspace is spanned by a single vector.  Later we generalize to
landscapes on $\reals^d$, in which case the perpendicular subspace has
dimension $d-1$.  The framework of Fig.~\ref{hillslope} is completed
by imposing fixed boundary conditions at the top of the slope (i.e.,
at $\xpar=0$), or by imposing the symmetry $\xpar\to-\xpar$.

Because Fig.~\ref{hillslope} explicitly distinguishes between the two
directions $\eperp$ and $\epar$, we expect this same anisotropy to be
reflected in the correlation function $C ({\bf r})$.  Thus, if $h$ is
self-affine, we expect different roughness exponents for correlations
measured in each of the directions $\epar$ and $\eperp$.  Specifically, we
define $\al_\Par$ and $\al_\Per$ such that $C_\Par(x_\Par) \sim
|x_\Par|^{\al_\Par}$ for correlations along fixed transects
$\vec{x}_\Per^0$ = const., and $C_\Per(\vec{x}_\Per) \sim
|\vec{x}_\Per|^{\al_\Per}$ for correlations along fixed transects
$x_\Par^0 =$ const., where in general $\al_\Par \neq \al_\Per$.  These
relations are summarized by the single scaling form
\begin{equation}
  C(x_\Par, \vec{x}_\Per)  \sim  b^{\al_\Par}  C(b^{-1} x_\Par,
  b^{-\zeta_\Par} \vec{x}_\Per),
  \label{scalaniso}
\end{equation}
where $\zeta_\Par$ is the {\em anisotropy exponent}. 
The roughness exponents $\al_\Par$  and $\al_\Per$ are related 
to $\zeta_\Par$ by 
\begin{equation}
  \al_\Per = \frac{\al_\Par}{\zeta_\Par}.
\label{zetapar}
\end{equation}
The anisotropy exponent $\zeta_\Par$ accounts for the different
rescaling factors along the two main directions. Since the space is
anisotropic, when performing a scale change, we must rescale $x_\Par$
and $\vec{x}_\Per$ by different factors $b_\Par$ and $b_\Per$,
respectively, if we are to recover a surface with the same statistical
properties.    
We assume that this scaling is self-affine, such that
\begin{equation}
  \zeta_\Par = \frac{\log b_\Per}{\log b_\Par} = \mbox{const}. 
\label{zaff}
\end{equation}
Note that $\zeta_\Par$ defines only the ratio of the roughness
exponents, but not their precise magnitudes; moreover, the scaling
form \equ{scalaniso} is not unique.  We can also express the rescaling
along the direction $\eperp$ by writing
\begin{equation}
  C(x_\Par, \vec{x}_\Per)  \sim  b^{\al_\Per}  C(b^{-\zeta_\Per} x_\Par,
  b^{-1} \vec{x}_\Per),
  \label{scalaniso2}
\end{equation}
where we have used the anisotropy exponent
$\zeta_\Per=1/\zeta_\Par$. Both scaling forms \equ{scalaniso} and
\equ{scalaniso2} are completely equivalent.

We seek a single stochastic equation for the landscape height $h$.  
We
assume that the deterministic motion of the underlying soil is locally
conserved such that
\begin{equation}
  \ptime{h} = - \nabla \cdot \vec{J} + \eta,
  \label{firstaniso}
\end{equation}
where $\vec{J}$ is the current of soil per unit length. The soil
however is not globally conserved, since it is lost at the bottom
boundary.  
Conservation is also broken by the
addition of a stochastic noise term $\eta$, discussed below.

Physically, the current $\vec{J}$ is expected to reflect two effects.
First, we expect a local isotropic diffusing component, tending to
smooth out the surface.  Second, we expect an average global flow of
dragged soil, directed mainly downhill.  Thus we postulate the
following form for the current:
\begin{equation}
  \vec{J} = -\nu \nabla h - \Gamma \ppar h.
  \label{soilcurrent}
\end{equation}
The first term corresponds to Fick's law for diffusion, and represents
the isotropic relaxational dynamics of the soil.  The second term
represents the average flow of soil that is dragged downhill, either
due to the flow of water or the scouring of the surface by the flow of
the soil itself.  The direction of this term is given by the vector $
\ppar h \equiv \partial_\Par h \: \epar $.  The term $\Gamma$ plays
the role of an {\em anomalous anisotropic diffusivity}.  In order to
gain insight into the role of $\Gamma$, consider the case in which
erosion results from the stress exerted on the soil bed by an overland
flow $Q$ of water, where $Q$ is the volumetric flow rate though unit
area perpendicular to the direction of steepest descent.  The greater
$Q$ is, the stronger is the stress \cite{iturbe97,scheidegger91}.
Moreover, since $Q$ flows downhill, it increases with distance
downslope. Thus $\Gamma$ must be an increasing function of $\xpar$.
Since the fixed inclination implies that $h$ increases with $\xpar$,
we choose to parameterize the anomalous diffusion as a function of the
height such that $\Gamma\equiv\Gamma(h)$.  Defining $\Gamma(h)=\la_0 +
g(h)$, with $g(0)=0$ and $G(h)=\int g(h) \,{\rm d} h$, we substitute
Eq.~\equ{soilcurrent} into \equ{firstaniso}.  Then, since
\begin{equation}
  g(h)\partial_\Par h = \frac{{\rm d} G(h)}{{\rm d} h} \partial_\Par h =
  \partial_\Par G(h),  
\end{equation}
where we have used the chain rule for the second equality, we obtain
\begin{equation}
  \ptime{h} = \nu_\Par \partial_\Par^2 h + \nu_\Per \nabla_\Per^2 h +
  \partial_\Par^2 G(h) + \eta,
  \label{interaniso}
\end{equation}
where $\nu_\Per=\nu$ and $\nu_\Par=\nu+\la_0$.

We may advance still further by making some additional assumptions.
Assuming that $\Gamma(h)$ is an analytical
function, we can perform a Taylor expansion in powers of $h$.  Since
all odd powers of $h$ must vanish in order to the preserve the joint
symmetry $h\to-h$, $\vec{J}\to-\vec{J}$ in Eq.~\equ{firstaniso}, we
are left at lowest order with $g(h) \simeq \la_2 h^2$. By dimensional
analysis (see next section and the Appendix) one can check
that all the terms $h^{2k}$ in this expansion are relevant under
rescaling.  However, the flux $Q(\xpar)$ of the erosive agent (water
or soil) flowing on the surface should grow no faster than $Q(\xpar)
\sim \xpar^{d}$.  
Then, taking $h \sim \xpar$, we find that the terms
in $g(h)$ should be of order $h^{d}$ or less.  Specializing to the
case of $d=2$ (i.e., real surfaces), we then find it reasonable to
truncate $g$ at second order. 
Equation \equ{interaniso} then takes the form
\begin{equation}
  \ptime{h} = \nu_\Par \partial_\Par^2 h + \nu_\Per \nabla_\Per^2 h +
  \frac{\lambda}{3}  \partial_\Par^2 (h^3) + \eta,
  \label{lastaniso}
\end{equation}
where $\la=\lambda_2$.  

Equation \equ{lastaniso} constitutes our full nonlinear theory.  Note
that it differs significantly from the anisotropic driven diffusion
equation of Hwa and Kardar \cite{hwa89,hwa92}.  The differences are
essentially due to the form of our current $\vec{J}$, which in our
case is suggested not only from symmetry principles, but also from the
physics of erosion.  We note additionally that, unlike some previous
models of landscape erosion that couple two equations---one for the
landscape $h$, and one for the overland flow $Q$
[41--44]%
---here we have
implicitly assumed that the effects of $Q$ may be subsumed into the
functional dependence of $\vec{J}$ on $h$.  Specifically, our
fundamental assumption of a preferred direction for the current
$\vec{J}$ can be traced to global constraints imposed by the fixed
inclination.  Thus {\em average} effects of $Q$---for example, the
fact that the erosion rate increases with $x_\parallel$ for
$\lambda_2>0$---survive our local formulation.

It remains to discuss the issue of noise.  We distinguish two possible
different sources.  First, we may allow a term of annealed
(time-dependent) noise, $\eta(\vec{x}, t)$, depending on time and
position, and describing a random external forcing due, for example,
to inhomogeneous rainfall.  We assume that this noise is isotropic,
Gaussian distributed, with zero mean, and correlations
\begin{equation}
  \left< \eta(\vec{x},t) \eta(\vec{x}',t') \right> = 2 D
  \delta^{(d)}(\vec{x}-\vec{x}') \delta(t-t').
  \label{eq:thermal-noise}
\end{equation}
Second, we may have a term of quenched (time-independent) noise to
account for the heterogeneity of the soil, mimicking the variations in
the erodibility of the landscape \cite{czirok93}.  The notion of
quenched noise is common in the study of interface growth in a
disordered medium, close to the depinning transition
\cite{amaral94}. In this case, the noise is a function of position
and height, $\eta(h, \vec{x})$, with correlations given in general by
\begin{equation}
  \left< \eta(h, \vec{x}) \eta(h', \vec{x}') \right> = 2 
  \Delta(h-h')  \delta^{(d)}(\vec{x}-\vec{x}') ,
\end{equation}
where the correlator 
$\Delta$ can be taken to be the usual Dirac delta function.  For this
prescription of quenched noise, however, an analytical approach 
appears hopeless
\cite{natterman92b,narayan93}.
We therefore propose to relax
this  definition and represent the quenched randomness of the soil by a
{\em static} noise  $\eta(\vec{x})$, with correlations
\begin{equation}
  \left<  \eta(\vec{x}) \eta(\vec{x}') \right> = 2 D
\delta^{(d)}(\vec{x}-\vec{x}').
  \label{eq:static-noise}
\end{equation}
This form of ``columnar'' noise, despite being a rather crude
approximation, has been 
previously proposed to model soil heterogeneity in cellular automata
models of fluvial networks \cite{caldarelli97}. 
Moreover, we have found it to be useful
for obtaining realistic river networks in
numerical simulations \cite{dodds98}.
In this paper we primarily consider the case of static noise
\equ{eq:static-noise}, corresponding to the limit in which the
external forcing is constant and the dominant source of noise is the
inhomogeneous composition of the soil.  Results for the case of
thermal noise are described in Ref.~\cite{pastor98}.

\section{Statistical solutions}

\subsection{Linear regime}

Even in the absence on any nonlinearity, fundamental conclusions may
be drawn from \equ{interaniso}. By setting $g=0$ (i.e., by considering
$\Gamma(h)=\la_0\equiv$ const.), we obtain the linear equation 
\begin{equation}
  \ptime{h} = \nu_\Par \partial_\Par^2 h + \nu_\Per \nabla_\Per^2 h +
  \eta, 
  \label{lineal}
\end{equation}
which is an anisotropic counterpart of the Edwards-Wilkinson equation
\cite{whittle62,edwards82}.  Let us consider for the moment the case
of a thermal source of noise $\eta$ with correlations
\equ{eq:thermal-noise}. In this case, it can be easily shown that the
amplitude of the correlation functions along the main directions
$\epar$ and $\eperp$ are inversely proportional to the square root of
the diffusivities $\nu_\Par$ and $\nu_\Per$, respectively. This
inverse proportionality is well known in the isotropic case (see
Ref.~\cite{natterman92}, Eqs.(2.19), (2.23), and (2.26)). In the
anisotropic case, we need only realize that the computation of,
say, $C_\Par$ follows from the correlations of $h$ computed at fixed
values of $\vec{x}_\perp$.  We then obtain $C_\Par \sim
\nu_\Par^{-1/2}$ and an analogous result for $C_\Per$.  Thus, in the
linear regime \equ{lineal}, the ratio of the correlations in the two
principal directions scales like
\begin{equation}
  \label{eq:relation}
  \frac{C_\Per}{C_\Par} \sim \left( \frac{\nu_\Par}{\nu_\Per}
  \right)^{1/2} .
\end{equation}
In other words, since the preferred
direction gives $\nu_\Par > \nu_\Per$ (since the relaxation of
material is expected to be faster in the direction $x_\Par$),
the topography is quantitatively rougher, at all scales and by the
same factor, in the perpendicular direction than in the parallel
direction.

In the case of static noise with correlations \equ{eq:static-noise},
it can be shown by dimensional analysis that the correlation 
functions are inversely proportional to the diffusivities.
In other words, $C \sim \nu^{-1}$
and Eq.~\equ{eq:relation} is correspondingly changed. The
qualitative prediction $C_\Per > C_\Par $, however,  
still holds.

\subsection{Nonlinear regime}

In this section we summarize the application of the dynamic
renormalization group (DRG) \cite{hwa92,ma75,forster77,medina89} to
our nonlinear model,
Eq.\ \equ{lastaniso}. Further details may be found in the Appendix.

The DRG proceeds in Fourier space by integrating out the fast modes,
corresponding to large momentum $\vec{k}$, over an outer shell
$\Lambda b^{-1} < k < \Lambda$, where $\Lambda$ is the upper cutoff in
wave vector space.  and $b$ is a rescaling factor. In order to bring
the system back to its original size, a rescaling is afterwards
performed, through the homogeneous transformation
\begin{equation}
  h(x_\Par, \vec{x}_\Per, t) = b^{\al_\Per} h(b^{-\zeta_\Per}
  x_\Par,   b^{-1} \vec{x}_\Per, b^{-z_\Per} t) .
  \label{rescal}
\end{equation}
The anisotropy is given by the exponent $\zeta_\Per \equiv
\zeta_\Par^{-1}$ [compare with Eqs.~\equ{scalaniso}
and~\equ{scalaniso2}], and the scaling with respect to time, discussed
below, is given by the {\em dynamic critical exponent} $z_\perp$.

After performing this transformation, we are left with an equation
with the same form as the original one, but with
different---renormalized---parameters. The transformation law of these
parameters under an infinitesimal rescaling $b=e^{{\rm d} \ell}$,
${\rm d} \ell \to 0$ constitutes the {\em flow equations} of the RG.
We are interested in the stable fixed points of these equations,
corresponding to scale invariant phases in the hydrodynamical limit.

To first order in the coupling constant $\la$, the flow equations read
\begin{eqnarray}
  \frac{{\rm d} \nu_\Par}{{\rm d} \ell} &=& \nu_\Par (z_\Per - 2
  \zeta_\Per +  \bar{\la}), \label{flow1}\\
  \frac{{\rm d} \nu_\Per}{{\rm d} \ell} &=& \nu_\Per (z_\Per - 2),
  \label{flow2}\\ 
  \frac{{\rm d} \la}{{\rm d} \ell} &=& \la \left(z_\Per + 2 \al_\Per - 2
  \zeta_\Per - \frac{3}{2} \bar{\la}\right),  \label{flow3}\\    
  \frac{{\rm d} D}{{\rm d} \ell} &=& D (2 z_\Per - 2 \al_\Per -
  \zeta_\Per - d + 1).   \label{flow4}
\end{eqnarray}
Here we have defined the effective coupling constant
\begin{equation}
  \bar{\la} = \frac{\la D}{2 \nu_\Par^{3/2}
    \nu_\Per^{3/2}} \Lambda^{d-4} K_{d-1},
  \label{effectivecoupling}
\end{equation}
with $K_d = S_d /(2\pi)^d$ and $S_d$ the surface area of a
$d$-dimensional unit hypersphere.  The flow equations for $\nu_\Per$
and $D$ are exact to all orders in the perturbation expansion. In the
case of $D$, this can be proven to be true for any stochastic equation
with a conserved current and a non-conserved noise, independently of
the details of the current \cite{lai91}.  For $\nu_\Per$, this results
from the fact that the nonlinearity is proportional to the external
momentum $k_\Par$, and cannot therefore renormalize a parameter
proportional to $k_\Per$ (see Eq.~\equ{lastaniso}) \cite{hwa92}. Thus
Eqs.~\equ{flow2} and \equ{flow4} provide us with the exact result
$z_\Per=2$, and with the exact scaling relation $2 z_\Per - 2 \al_\Per
- \zeta_\Per = d - 1$. 
The dynamic critical exponent $z_\Per$ measures the
saturation length of correlations as a function of time
\cite{sornette93,barabasi95}. 
However, since the time scales for geomorphologic 
evolution are many orders of magnitude larger than those 
available for observation, the actual value of $z_\Per$,
appears, at least at this point, to be purely a matter of speculation.

Given \equ{flow1}-\equ{effectivecoupling}, the effective coupling
$\bar{\la}$ flows under rescaling as
\begin{equation}
  \frac{{\rm d} \bar{\la}}{{\rm d} \ell} =\bar{\la} (\ep - 3 \bar{\la}),
  \label{floweq}
\end{equation}
where $\ep=4-d$ . The stable fixed points of \equ{floweq} are
$\bar{\la}^*_1 = 0$ for $d>4$ and $\bar{\la}^*_2 = \ep / 3$ for $d<4$.
For $d>4$ the critical exponents attain their mean-field values
$\al_\Per^{0}=0$, $\zeta_\Per^{0} = 1$, and $z_\Per^{0} = 2$.
On the other hand, for $d<4$, the critical exponents
computed at first order in $\ep$ are
\begin{equation}
  \al_\Per=\frac{5 \ep}{12}, \qquad \zeta_\Per = 1 + \frac{\ep}{6},
  \qquad z_\Per =2 .
  \label{exponents}
\end{equation}

The physically relevant dimension for erosion in the real world is
$d=2$.  Even though the result~\equ{exponents}  represents only the
first terms in a expansion in powers of $\ep$, and is therefore
valid only approximately for small values of that variable, 
we can still gain some
information by setting $\ep=2$. In this case, we obtain the scaling
exponents 
\begin{equation}
  \al_\Per = \frac{5}{6} \simeq 0.83, \:
  \zeta_\Per = \frac{1}{\zeta_\Par} = \frac{4}{3},  \:
  \al_\Par=\frac{\al_\Per}{\zeta_\Per} = \frac{5}{8} \simeq 0.63.  
  \label{numpredict}
\end{equation}
\section{Comparison with field data}

The values \equ{numpredict} predicted for $\al_\Per$ and $\al_\Par$
are in reasonable agreement with previous measures made at small
length scales
[41--48],
where the roughness exponent was reported to be between $0.70$ and
$0.85$. This observation lends support to the hypothesis of two
``universality classes'' for geomorphological evolution. The first of
these classes encompasses topographies evolving isotropically at large
length scales, and possibly dominated by non-erosive mechanisms such
as internal tectonic stresses.  A characteristic of this class is a
small roughness exponent which is compatible with estimates made from
the KPZ equation \refeq{kpz}.  Thus, following the proposal of
Sornette and Zhang, we can identify the KPZ equation as a description
of some universal features of the large scale dynamics.  On the other
hand, for small length scales we expect anisotropic effects to be
dominant. The anisotropy, which is induced by small-scale inclinations
of the landscape, would lead to a purely erosive dynamics, which in
turn should yield the large roughness exponents predicted by our
theory.

\begin{figure}[t]
  \centerline{\epsfig{file=\FigPSPath/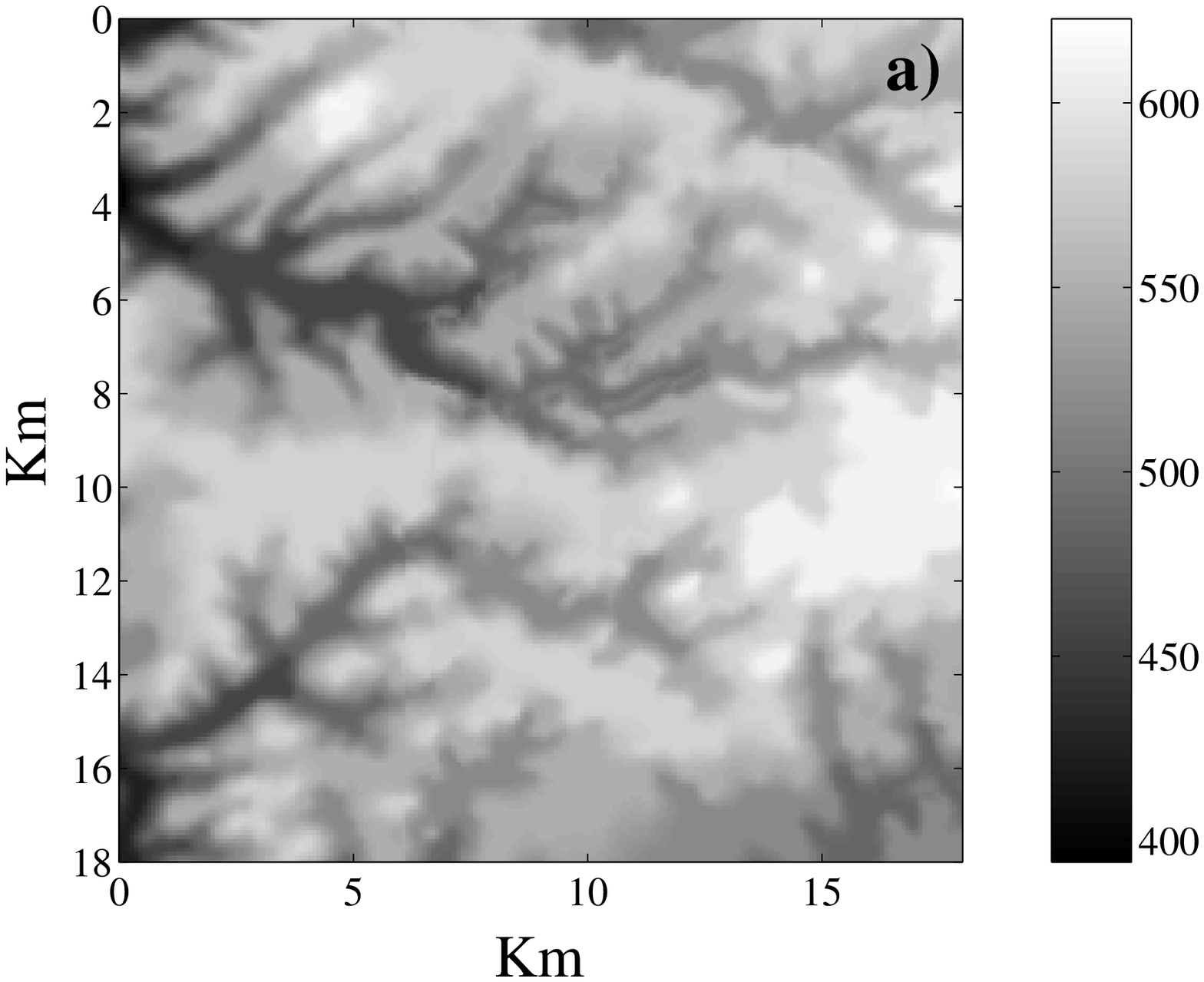, width=6truecm,
      bbllx=106, bblly=228, bburx=458, bbury=580}
    \hspace*{3cm}\epsfig{file=\FigPSPath/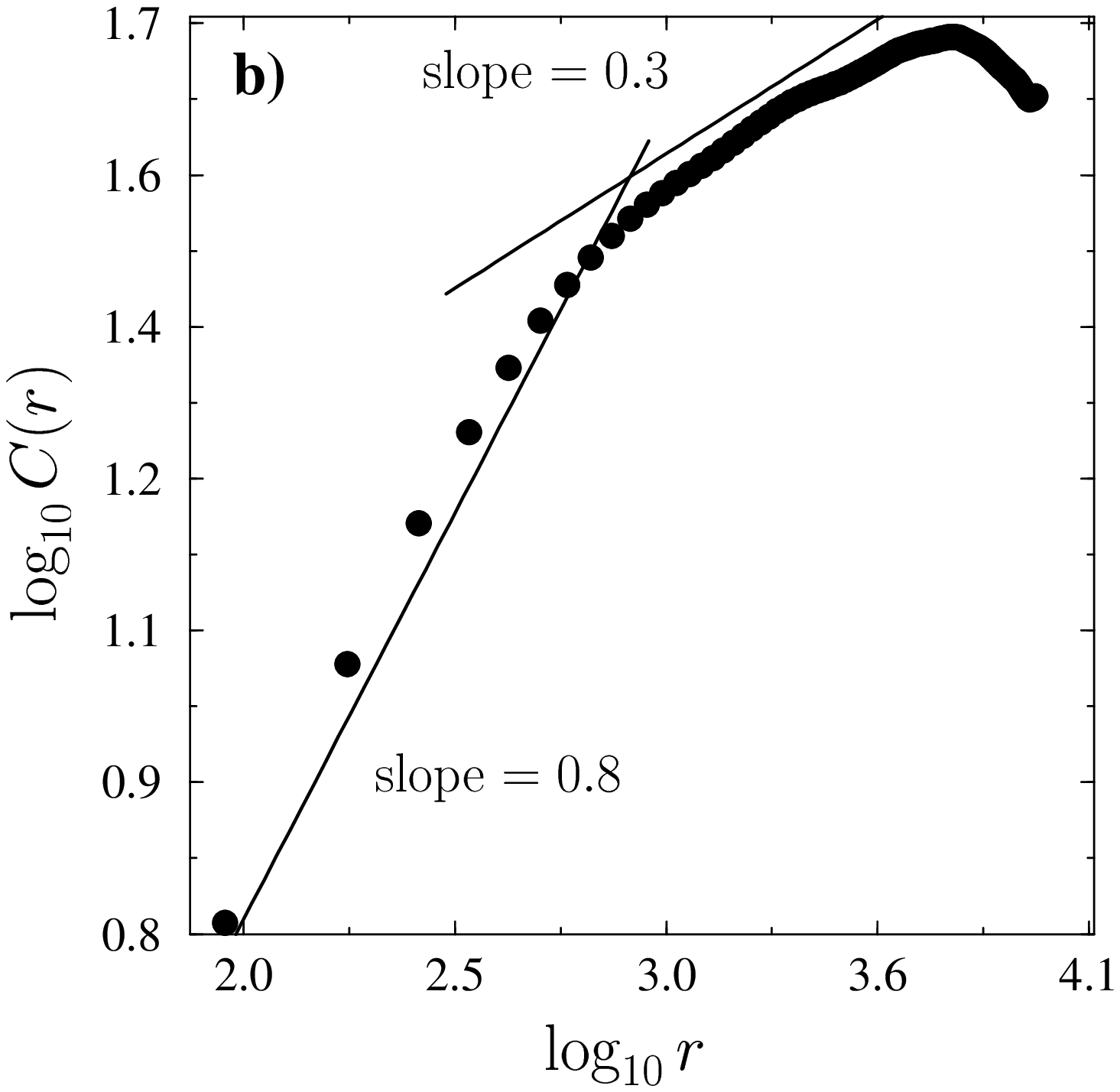,
      width=6truecm, bbllx=212, bblly=317, bburx=553, bbury=662}}
  \vspace*{1.1cm}
  \caption{a) Digital elevation map of an area of the Appalachian
    Plateau, in Northwest Pennsylvania. Elevations are given in
    meters.  The spatial resolution is 90 m.  b) Averaged
    height-height correlation function $C(r)$ for the landscape in
    Figure~\protect\ref{warren}a, where $r$ is oriented in the
    vertical direction of (a).  A plot of similar shape, but with
    smaller values of $C(r)$, is obtained in the horizontal case.
    Logarithms are computed from quantities measured in units of
    meters.}
  \label{warren}
\end{figure}

If these two universality classes do indeed exist at different length
scales, then one should be able to find evidence of a crossover from
one regime to another in the same piece of topography.  Specifically,
one expects that the correlation function should change from a high
$\al$ regime at short length scales to a low $\al$ regime at large
length scales.  This sort of crossover has indeed been reported
several times in the literature
\cite{mark84,matsushita89,chase92,lifton92,dietler92}.  Indeed,
Ref.~\cite{matsushita89} has already suggested that the crossover
length separates a small-scale erosive regime from large-scale
tectonic deformation.  Ref.~\cite{dietler92}, on the other hand,
suggests that the crossover separates length scales which have had
sufficient time to fully develop in a KPZ-like way and those which
have not.  Although it is beyond the scope of this paper to make a
definitive argument in favor of either of these interpretations, in
Figure~\ref{warren} we present measurements of our own, from the
Appalachian Plateau in NW Pennsylvania.  Figure \ref{warren} also
suggests the presence of a crossover, here located at a characteristic
scale of about 1 km.  We note that in topography depicted in Figure
\ref{warren}a, the principal features are deeply eroded channels with
a characteristic width of order 1 km.  The long wavelength features,
on the other hand, have resulted from tectonic stresses associated
with the formation of the Appalachian Mountains.  Thus, based on the
evidence of this example, we prefer the interpretation of
Ref.~\cite{matsushita89}.

\begin{figure}[t]
  \centerline{\epsfig{file=\FigPSPath/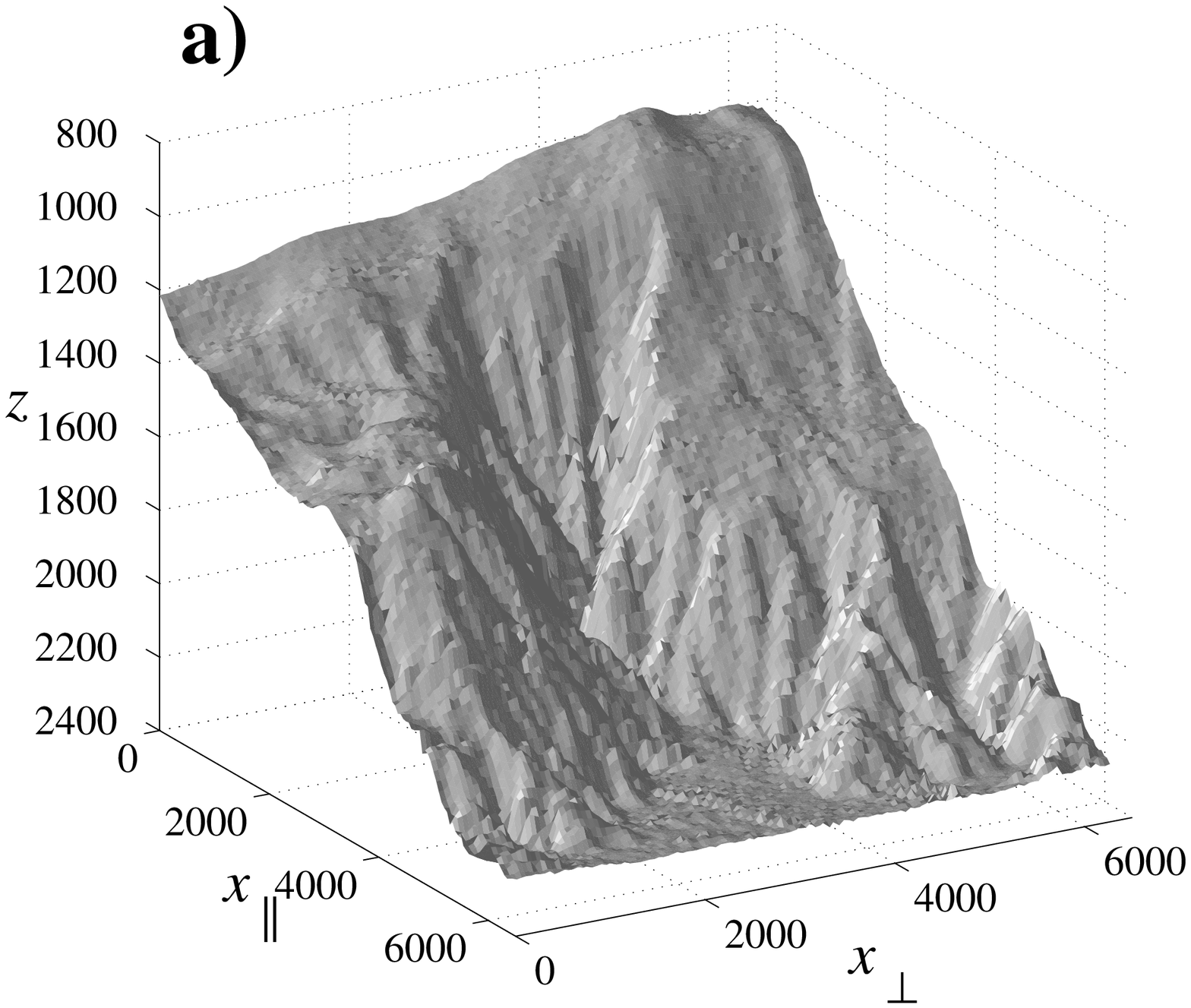, width=5.5truecm,
      bbllx=93, bblly=194, bburx=536, bbury=615}
    \hspace*{2.5cm}\epsfig{file=\FigPSPath/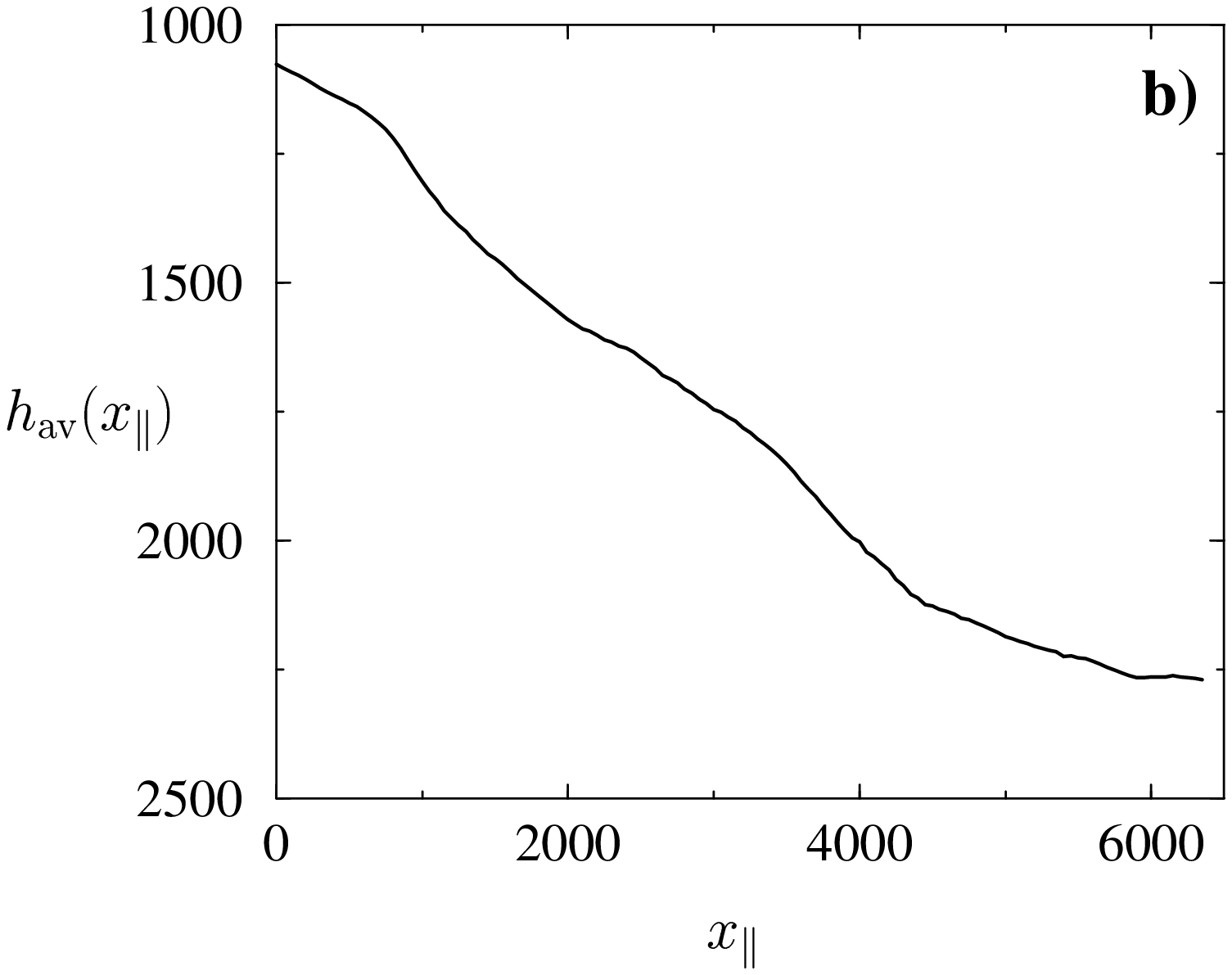, width=6truecm,
      bbllx=213, bblly=385, bburx=546, bbury=657}} \vspace*{1cm}
  \caption{a) Digital map of a submarine canyon off the coast of
    Oregon, located at coordinates $44^o 40'$ N, $125^o 45'$ W. The
    vertical axis represents the depth $z$ below sea level.  The
    spatial resolution is 50 m.  b) Mean average profile of the
    canyon, along the direction $x_\Par$.  All units are given in
    meters.}
  \label{submarine}
\end{figure}

In Figure 2 there is no obvious preferred direction,
and all of the measurements reported in the literature were 
either averaged over all directions or the direction of the 
measurements was not reported.
Thus, to check the full validity of our results with natural
topography that has an unambiguous preferred direction, we have
analyzed digital bathymetric maps of the continental slope off the
coast of Oregon.  In this case the slope results from the relatively
abrupt increase in the depth of the seafloor as the continental shelf
gives way to the deeper continental rise.  Figure~\ref{submarine}a
shows one portion of this region.  Here the main feature of the
topography is a deep incision called a {\em submarine canyon}.  In
this region, submarine canyons are thought to have resulted from
seepage-induced slope failure \cite{orange94}, which occurs when
excess pore pressure within the material overcomes the gravitational
and friction forces on the surface of the material, causing the slope
to become unstable.  Slope instabilities then create submarine
avalanches, which themselves can erode the slope as they slide
downwards.

\begin{figure}[t]
  \centerline{\epsfig{file=\FigPSPath/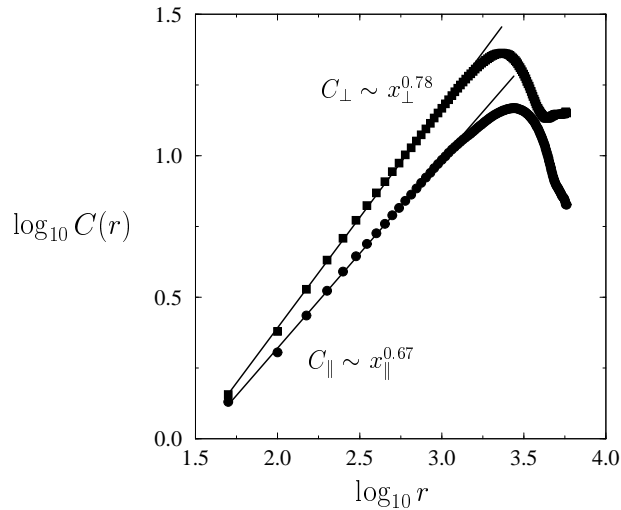, width=8.5cm,
      bbllx=35, bblly=245, bburx=555, bbury=680}}
  \caption{Height-height correlation functions computed 
    along the parallel ($C_\Par$) and perpendicular ($C_\Per$)
    directions for the landscape shown in
    Fig.~\protect\ref{submarine}a. Solid lines are least-squares fits
    to the scaling region.  The logarithms are computed from
    quantities measured in meters.}
  \label{correls}
\end{figure}

In order to make comparisons with our predictions \equ{numpredict}, we
have computed the correlation functions $C_\Par$ and $C_\Per$,
corresponding, respectively, to the parallel and perpendicular
directions of the seafloor topography in Fig.~\ref{submarine}a. The
computation of $C_\Per$ follows naturally from its definition but the
computation of $C_\Par$ deserves some comment. The fluctuations
measured by $C_\Par$ must be defined with respect to an appropriate
average profile. One expects that geologic processes other than
erosion (e.g., tectonic stresses) are responsible for long-wavelength
deformations in the parallel direction. Assuming that these
deformations are on average constant in the perpendicular direction,
we may estimate such systematic corrections by computing the mean
profile along the parallel direction,
\begin{equation}
  h_{\rm av} (x_\Par) = \frac{1}{L_\Per} \int {\rm d} x_\Per  h
  (x_\Par, x_\Per), 
\end{equation}
where $L_\Per$ is the length of the system in the perpendicular
direction. We have plotted $h_{\rm av}$ in Fig.~\ref{submarine}b. We
use it to detrend $h$ by computing the correlation function $C_\Par$
from the fluctuations of the detrended surface $\tilde{h} = h-h_{\rm
  av} (x_\Par)$.

Figure~\ref{correls} shows the plots of $C_\Par$ and $C_\Per$,
corresponding to the topography in Fig.~\ref{submarine}a.  One sees
that the least-squares estimates of the roughness exponents, $\al_\Par
\simeq 0.67$ and $\al_\Per \simeq 0.78$, exhibit a surprisingly good
fit to our theoretical predictions \equ{numpredict}.

\begin{figure}[t]
  \centerline{\epsfig{file=\FigPSPath/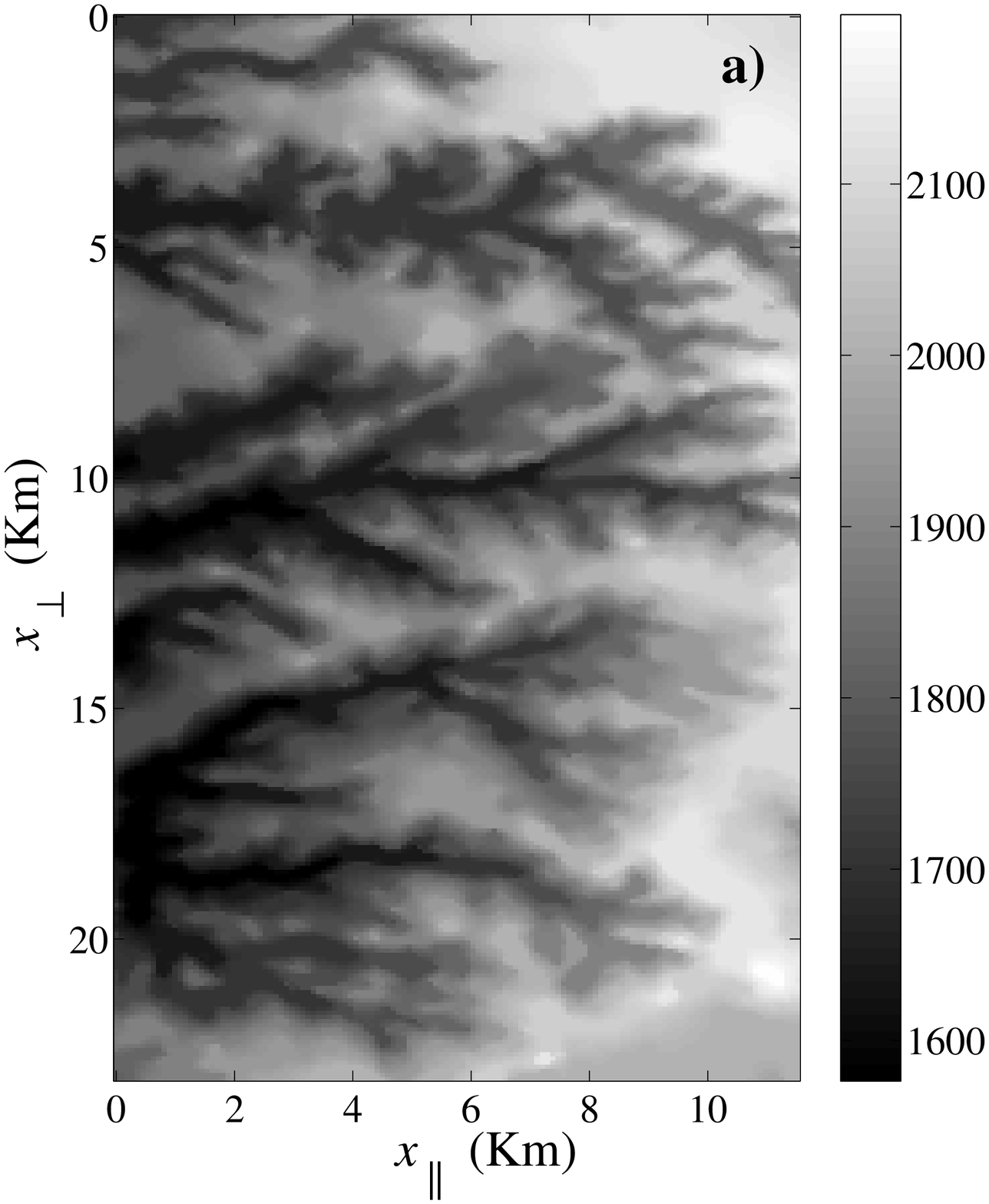, width=4.7truecm,
      bbllx=91, bblly=112, bburx=473, bbury=707}
  \hspace*{3cm}\epsfig{file=\FigPSPath/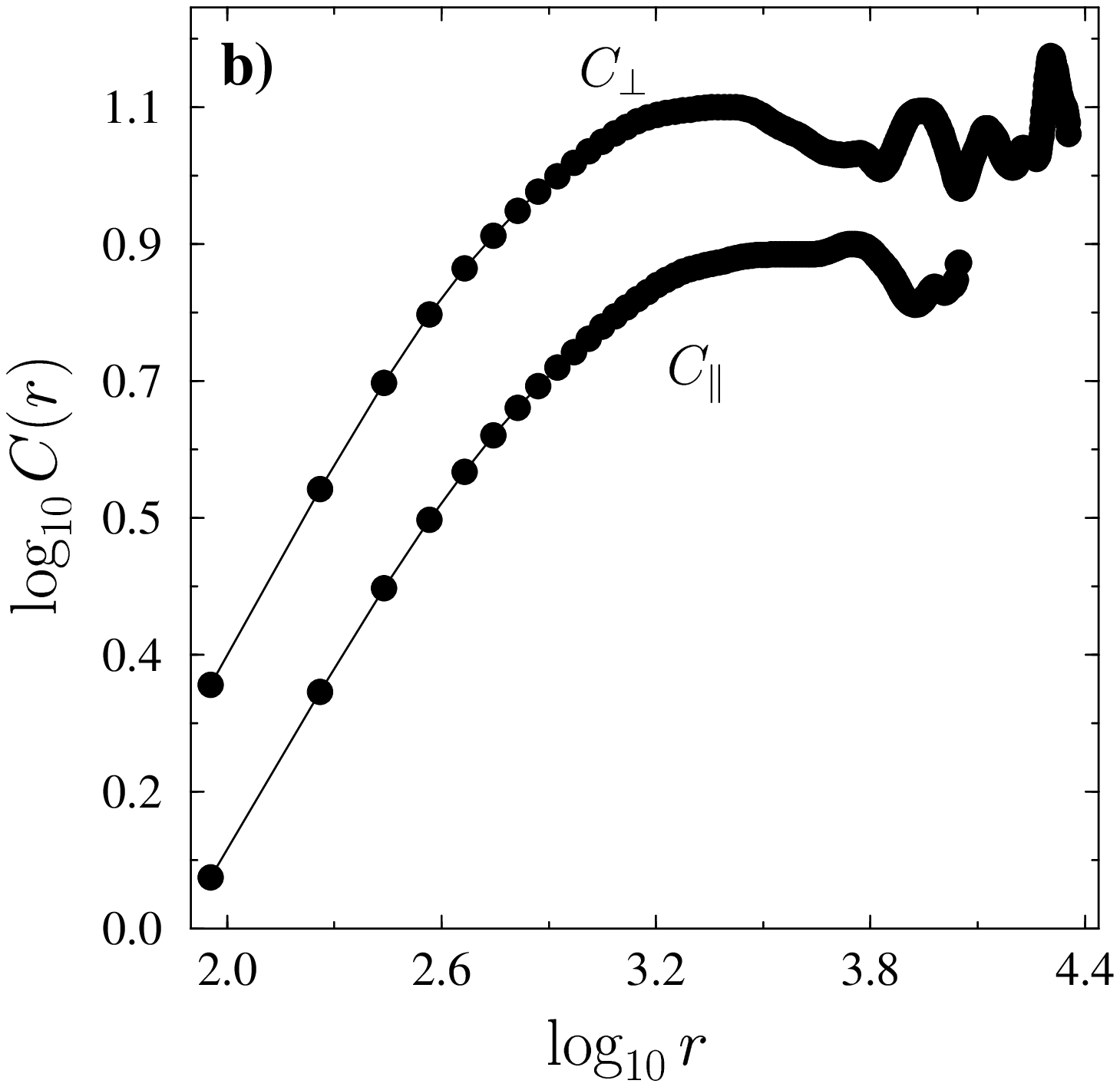, width=7truecm,
    bbllx=212, bblly=324, bburx=552, bbury=669}}
 \vspace*{1cm}
  \caption{a) Digital elevation map of an area near Marble
    Canyon, in Northeast Arizona. Elevations are given in meters, and
    the spatial resolution is 90 m.  b) Height-height correlation
    functions computed along the parallel ($C_\Par$) and perpendicular
    ($C_\Per$) directions for the landscape shown in
    Fig.~\protect\ref{marble}a.  Logarithms are computed from
    quantities measured in meters.}
  \label{marble}
\end{figure}

We have also measured $C_\Par$ and $C_\Per$ in some desert
environments. One such example is shown in Figure~\ref{marble},
corresponding to an area near Marble Canyon, in Northeast Arizona.  In
these cases we do not obtain conclusive power law scaling, but we almost
always find $C_\Per / C_\Par > 1$, as predicted by the linear theory.
(Indeed, at small scales in this case, we find $C_\Per \simeq 1.8 C_\Par$)
Thus, while the example of
Figure~\ref{submarine} may be in some sense specialized, one of our
main predictions---that the topography in the perpendicular direction
is rougher than the topography in the parallel direction---seems to be
of fairly general validity.

\section{Size distribution of avalanches}

Real sloping topography can erode episodically in a series
of infrequent events.  This episodic erosion amounts to a series
of {\em avalanches}.

For the case of sloping submarine topography such as that shown in
Figure \ref{submarine}a, such avalanches can create gravity-driven
flows \cite{simpson97} of suspended sediment.  When these flows
finally come to rest, the sediment settles out.  Then, over geologic
time, the sediment solidifies to form sedimentary rocks known as {\em
  turbidites} \cite{press82}.  Partly because these sedimentation
events are widely spaced in time, individual layers of rock may be
associated with each avalanche-like sedimentation event.  The
thickness of these layers may be assumed to be related to the size, or
volume of sediment, associated with the avalanche that created them.

Recent empirical studies of turbidite deposits show that in some
instances a power-law distribution of thicknesses may be observed that
extends over nearly two orders of magnitude in thickness
\cite{hiscott92,rothman94,rothman95}.  In other words, the
measurements indicate that the probability $P_{\Delta} (\Delta)$ of an
avalanche resulting in a deposit of thickness $\Delta$ scales like
\begin{equation}
  P_\Delta ( \Delta ) \sim \Delta^{-\gamma},
  \label{eq:turbidite}
\end{equation}
where $\gamma$ is a characteristic exponent.\footnote {Note that our
  notation differs from Refs.~\cite{rothman94,rothman95}, where the
  cumulative distribution of layers was studied.  Specifically, their
  exponent $B$ is equal to, in our notation, $\gamma -1$.}  
In the best documented
cases, $\gamma$ is between about $2$ and $2.4$
\cite{hiscott92,rothman94,rothman95}.

Refs.~\cite{rothman94,rothman95} suggested that the power-law
distribution \refeq{eq:turbidite} could be the result of a natural
manifestation of {\em self-organized criticality} (SOC) \cite{jensen98,bak88}.
Systems exhibiting SOC, and in particular, certain models of
sandpiles, exhibit a dynamics dominated by avalanche events, in which
the number of avalanches of size $s$ scales like the power law
\begin{equation}
  P_s (s) \sim s^{-\tau},
  \label{equ:powerlaw}
\end{equation}
with a characteristic exponent $\tau$. 
Assuming that turbidites result from a series of slumps that may be in
some way related to the SOC sandpile models,
then, as indicated in Refs.~\cite{hiscott92,rothman94,rothman95},
their size distribution may also be given according to 
a expression similar to \equ{equ:powerlaw}. 
One might further expect that this distribution
could be related to geometric aspects of the surface from which the
avalanches fall.
Our objective in this section, then, is to relate the scaling
properties of the topography of a sloping surface to the
scaling properties of the avalanche size distribution.

\begin{figure}[t]
  \centerline{\epsfig{file=\FigPSPath/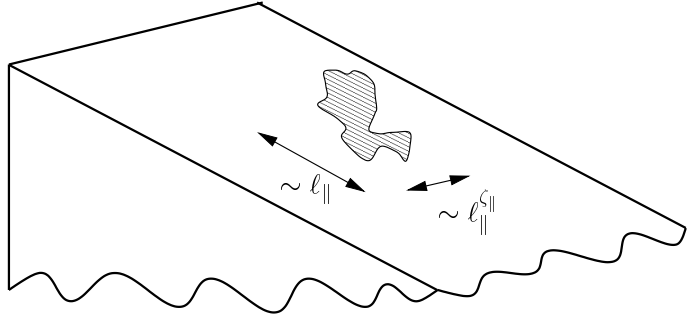, width=10truecm}}
  \caption{Scaling of an ``avalanche patch'' over an anisotropic
    landscape.}
  \label{avalanche}
\end{figure}

To do so, we follow Refs.~\cite{hwa89,hwa92} and view our model
\equ{lastaniso} as a transport equation that describes a driven
diffusive system, i.e., a sandpile.  Under this assumption, we may
compute the probability distribution of the size of the surface area
that relaxes as a result of the interplay between the driving
force---noise---and the diffusive damping.  We assume that this
distribution is a power law, with a cutoff due to finite-size
effects.  Moreover, we expect these relaxing surface patches to
exhibit the same anisotropy as the underlying topography.  In other
words, as shown in Figure \ref{avalanche}, we assume that the
avalanches result from unstable patches of unit thickness with
extension proportional to $\ell_\parallel$ and $\ell_\perp$ in the
parallel and perpendicular directions, respectively.  From equations
\refeq{scalaniso}, \refeq{zetapar}, and \refeq{zaff}, the self-affine
nature of the topography gives $\ell_\perp \sim
\ell_\parallel^{\zeta_\parallel}$.  Thus the size $s$ of an unstable
patch of extent $\ell_\parallel$ in the parallel direction scales like
$\ell_\parallel^{1+\zeta_\parallel}$, and the maximum size of an
avalanche in a system of parallel extent $L$ scales like
$L^{1+\zeta_\parallel}$.  The probability distribution of avalanche
sizes $s$ in a system of finite size $L$ may then be expressed as
\cite{bak88}
\begin{equation}
  P(s, L) = s^{-\tau} f\left( \frac{s}{L^{1+\zeta_\parallel}} \right) .
  \label{equ:anisoscal}
\end{equation}
Here $f(x)$ is a scaling function such that $f$ is constant for small
$x < 1$ and zero for large $x > 1$.  Using equation
\refeq{equ:anisoscal}, we can relate $\tau$ to $\zeta_\parallel$ by
means of a scaling argument \cite{caldarelli97,amaral96,tadic98}.  The
average size $\langle s \rangle$ of an avalanche is defined by
\begin{equation}
  \langle s \rangle = \int s^{1-\tau}
  f\left( \frac{s}{L^{1+\zeta_\parallel}} \right) {\rm d}s .
\end{equation}
Performing the change of variables $s = \xi L^{1+\zeta_\parallel}$
then yields
\begin{eqnarray}
  \left< s \right> &=& L^{(2-\tau)(1+\zeta_\parallel)} \int \xi^{1-\tau}
  f\left( \xi  \right) {\rm d}\xi \\ 
  & \sim &
  L^{(2-\tau)(1+\zeta_\parallel)}.
  \label{equ:exporel}
\end{eqnarray}
On the other hand, since each avalanche is a self-affine patch,
as $L \rightarrow \infty$ the average patch should become 
increasingly elongated in the parallel direction
(since $\zeta_\parallel < 1$).
Thus one expects
$\langle s \rangle \sim L$ for large $L$.
This relation,
together with \equ{equ:exporel}, provides us with the result
\begin{equation}
  \tau = 2 - \frac{1}{1+\zeta_\parallel},
  \label{equ:secondexporel}
\end{equation}
which relates the avalanche-size exponent $\tau$ to the anisotropy
exponent $\zeta_\parallel$.

As noted in Ref.~\cite{rothman95,malinverno97}, we must also address
the relationship between the avalanche size (or mass) $s$ and the thickness
$\Delta$ of the deposited layer.  One way to do this is by introducing
the {\em spreading exponent} $\chi$ such that
\begin{equation}
  \Delta \sim s^\chi.
  \label{equ:spreading}
\end{equation}
For perfect spreading (i.e., all layers have the same thickness),
$\chi=0$, whereas for no spreading at all (i.e., all sedimentation
events cover the same area $A=s/\Delta = $ const.), $\chi=1$.
Empirical studies of rock slides, for example, indicate that $\chi
\sim 1/3$ \cite{davies82}, corresponding to {\em self-similar
  areal spreading}.  From the relation $P_s (s) = P_\Delta ( [ \Delta
( s ) ] ) ( \mbox{d} \Delta / \mbox{d} s )$, we find, using equations
\equ{equ:anisoscal} and \equ{equ:spreading}, that
\begin{equation}
  \gamma = 1 + \frac{\tau -1}{\chi}.
\end{equation}
Then from Eq.~\equ{equ:secondexporel} we obtain 
\begin{equation}
\label{socdr}
  \gamma = 1 + \frac{1}{\chi}
\left (
\frac{\zeta_\parallel}{1+\zeta_\parallel}
\right ) .
\end{equation}

Equation \refeq{socdr} relates the exponents describing surface
anisotropy, avalanche size, and spreading.  Using our DRG estimate for
the anisotropy exponent, $\zeta_\parallel=3/4$, and assuming a
spreading exponent $\chi=1/3$, we find that
\begin{equation}
  \gamma = \frac{16}{7} \approx 2.3,
\end{equation}
which is in reasonable agreement with the best documented results
of Refs.~\cite{hiscott92,rothman94,rothman95}.
On the other hand, we may consider equation \refeq{socdr} to be
a prediction of the spreading exponent $\chi$ when $\gamma$ is measured
in a single location but $\chi$ is unavailable.
This could be useful in geological applications where one wishes
to know the spatial extent of a sequence of turbidite deposits.

\section{Conclusions}

In concluding, it is worthwhile to reflect on the main elements of our
theory.  Lacking any fundamental ``equations of motion'' for erosion,
we have elected to proceed from conservation laws and symmetry
principles.  Thus the principal ingredients of our model are the
conservation of the eroding material, the presence of a preferred
direction for the transport of it, and randomness in either the
landscape or the forcing.  Making just these assumptions, we derived
an anisotropic stochastic equation from which we have extracted both
qualitative and quantitative predictions.  The main qualitative
prediction is that eroded topography is rougher in the direction
across slopes than it is in the direction down slopes.  The main
quantitative predictions are scaling laws for height-height
correlation functions.  These require additional assumptions or
restictions concerning the noise and the relevant degree of
nonlinearity.  Both the qualitative and quantitative predictions
appear to be in good agreement with measurements made from real
landscapes.

We have also included an interpretation of our theoretical model 
as a driven diffusion equation.
In this case we have been able to relate the distribution of the
sizes of erosion events, or ``avalanches,'' to the self-affine scaling
that we have predicted for the nonlinear regime of our model.
The testing of this prediction is beyond the scope of this paper, however,
as it would require either unusually extensive geological data
or an innovative laboratory experiment.

Our results apply, in principle, to any erosive
process with the appropriate lack of symmetry. 
In the usual geological setting, however, 
the anisotropy applies specifically to a surface of
fixed inclination which, in turn, implies that our theory should only
apply locally, to the relatively small scales where the preferred
direction of transport is approximately constant. 
Because the anisotropy should vanish at large length scales, 
we argue that large-scale features of topography
should be presumably described by a different, isotropic theory,
such as the KPZ equation \refeq{kpz} \cite{sornette93,kardar86}. 
We provide evidence, and cite additional results from the literature,
that such a crossover indeed exists.
We suggest, in line with others \cite{matsushita89},
that the crossover length separates small-scale, externally induced,
erosive features of landscapes from large-scale deformations 
(such as those induced by tectonic stresses) of internal origin.

Finally, we wish to note that, while there is much evidence that
landscapes can be self-affine, this evidence
is rarely unambiguous and certainly not ubiquitous.
Our qualitative prediction that $C_\Per > C_\Par$,
on the other hand, appears more robust than any predictions
of scaling exponents, or even scaling itself.
Our results suggest that the coupling of anisotropy to
topographic orientation may be a fundamental physical property 
of eroding landscapes.
Precisely which length scales are relevant to this coupling,
however, remains an open question.

\section*{Acknowledgments}

We thank B. Tadi\'{c} for fruitful discussions and suggestions. R.P.S.
acknowledges financial support from the Ministerio de Educaci\'{o}n y
Cultura (Spain). The work of D.H.R. was partially supported by NSF
grant EAR-9706220.

\newpage

\section*{Appendix}

In this Appendix we develop further details of the renormalization
group analysis of Eq.~\equ{lastaniso}, where $\eta$ is a static noise
term, Gaussian distributed, with zero average and correlations
\begin{equation}
  \left< \eta(\vec{x}) \eta(\vec{x}') \right> = 2 D
  \delta^{(d)}(\vec{x}-\vec{x}').
\end{equation}
Here $D$ is a parameter gauging the strength of the noise. 

\begin{figure}[t]
  \centerline{\epsfig{file=\FigPSPath/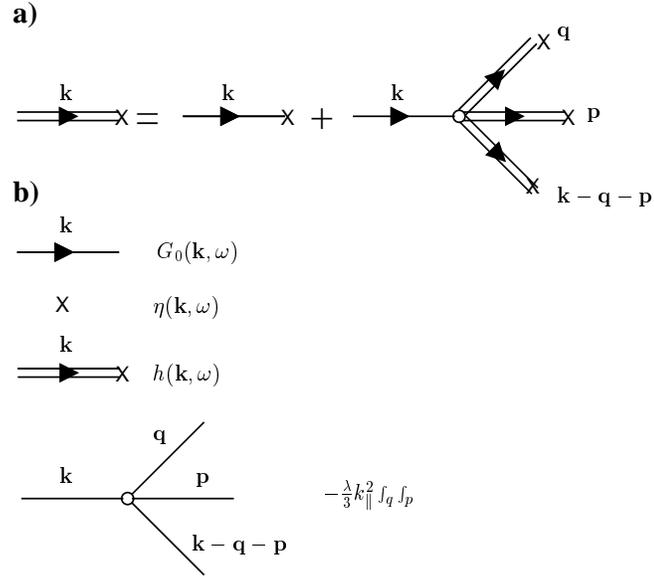, width=9cm,
      bbllx=137, bblly=285, bburx=567, bbury=680}\protect\label{feynman}} 
  \caption{a)
    Diagrammatic expansion of Eq.~\equ{iterate} in Feynman diagrams.
    b) Definition of the various terms.}
  
\end{figure}

In order to proceed, we first Fourier transform the function $h$,
defining 
\begin{equation}
  h(\vec{x}, t) = \int_k  h(\vec{k}, \omega)  e^{i(\vec{k}\cdot\vec{x}
  - \omega t)},
\end{equation}
where we have defined
\begin{displaymath}
  \int_k \equiv  \int_{|\vec{k}|<\Lambda} \frac{{\rm
  d}^d\vec{k}}{(2\pi)^d} \int_{-\infty}^{\infty}   \frac{{\rm
  d}\omega}{2\pi}.
\end{displaymath}
The integrals over $\vec{k}$ are restricted to the upper cutoff
$\Lambda$, which  plays the role of a lattice spacing, or minimum
distance in real space.  In momentum space, and after performing a few
algebraic manipulations, Eq.~\equ{lastaniso} reads
\begin{equation}
  h(\vec{k}, \omega) = G_0(\vec{k}, \omega) \eta(\vec{k},\omega) -
  \frac{\la}{3} k_{\Par}^2 G_0(\vec{k}, \omega) \int_q \int_{q'}
  h(\vec{q}, \Omega)   h(\vec{q}', \Omega')
  h(\vec{k}-\vec{q}-\vec{q}', \omega 
  -\Omega-\Omega' ),
  \label{iterate}
\end{equation}
where $\eta(\vec{k}, \omega)$ is the Fourier transform of the static
noise $\eta(\vec{x})$, with correlations
\begin{displaymath}
  \left< \eta(\vec{k},\omega) \eta(\vec{q},\Omega) \right> = 2 D
  (2\pi)^{d+2} 
  \delta^{(d)}(\vec{k}+\vec{q}) \delta(\omega)  \delta(\Omega),
\end{displaymath}
and the free propagator $G_0(\vec{k}, \omega)$ has the form
\begin{equation}
  G_0(\vec{k}, \omega)=\frac{1}{\nupar k_\Par^2 + \nuperp k_\Per^2 -i
    \omega}.
  \label{freeprop}
\end{equation}
In Figure~\ref{feynman} we have represented Eq.~\equ{iterate}
in terms of Feynman diagrams \cite{medina89}.

As we mentioned above, the RG procedure consists of the elimination of
the fast modes (large wave vectors $\vec{k}$), followed by a rescaling
of the system back to its original size by means of the
transformation
\begin{displaymath}
  h(x_\Par, \vec{x}_\Per, t) = b^{\al_\Per} h(b^{-\zeta_\Per}
  x_\Par,   b^{-1} \vec{x}_\Per, b^{-z_\Per} t).
\end{displaymath}
The relevance of the different parameters of the problem ($\nu_\Par$,
$\nu_\Per$, $\la$, and $D$) can be estimated by dimensional analysis.
One can check that, under the aforementioned transformation, these
parameters rescale as
\begin{equation}
  \nupar'=b^{-2\zeta_\Per+z_\Per}\nupar, \quad \nuperp'=
  b^{-2+z_\Per} \nuperp, \quad \la'=b^{2\al_\Per -2\zeta_\Per + 
  z_\Per} \la, \quad D'=b^{-2\al_\Per + 2z_\Per - \zeta_\Per -(d
  -1)}D. 
  \label{eq:dimensional}
\end{equation}
The coupling constant $\la$ will be irrelevant\footnote{In the
  renormalization group framework, a parameter is irrelevant if it
  flows to zero, and can therefore be neglected in the calculations.}
when its scaling exponent, $\la'=b^{y_\la} \la$, is
negative, where $y_\la=2\al_\Per-2\zeta_\Per + z_\Per$. Selecting the
values of $z_\Per$, $\al_\Per$, and $\zeta_\Per$ so that the
transformations of $\nupar$, $\nuperp$, and $D$ are invariant, we
obtain
\begin{equation}
  z_\Per^0=2, \qquad \zeta_\Per^0=1, \quad \mbox{and} \quad
  \al_\Per^0=(4-d)/2,
  \label{meanfield}
\end{equation}
and from them, $y_\la=4-d$.  The nonlinearity is irrelevant (that is,
$y_\la<0$) when $d>d_c=4$. This defines the {\em critical dimension}
$d_c$ of our system. Above the critical dimension, the nonlinearity is
negligible, and we recover the scaling exponents \equ{meanfield},
which are simply given by dimensional analysis. On the other hand,
below $d_c$ the nonlinearity prevails, and we expect
fluctuations to be dominant and produce nontrivial scaling exponents.

The RG program is carried out with the help of diagrammatic
techniques \cite{barabasi95,ma75,medina89}. The idea is to graphically
iterate \equ{iterate} to the desired order in $\la$ and express the
equation in terms of {\em effective} parameters, which are given
as integrals of powers $G_0$.  Afterwards, integration of fast modes,
by averaging over the noise in the outer shell, and rescaling provides
us with the renormalized parameters and the consequent flow equations,
\equ{flow1}-\equ{flow4}, in the limit of an infinitesimal
transformation.

\end{document}